\documentclass[twocolumn]{jpsj3} 
\usepackage{bm}

\title{Origin of $T_c$ Enhancement Induced by Doping Yttrium and Hydrogen into LaFeAsO-based Superconductors: $^{57}$Fe-, $^{75}$As-, $^{139}$La-, and $^{1}$H-NMR Studies}

\author{
Hiroki Yamashita$^{1}$, Hidekazu Mukuda$^{1,3}$\thanks{E-mail address: mukuda@mp.es.osaka-u.ac.jp}, Mitsuharu Yashima$^{1,3}$, Satoshi Furukawa$^{1}$, Yoshio Kitaoka$^{1}$, Kiichi Miyazawa$^{2}$, Parasharam M. Shirage$^{2}$, Hiroshi Eisaki$^{2,3}$, and Akira Iyo$^{2,3}$ 
}

\inst{
$^{1}$Graduate School of Engineering Science, Osaka University, Toyonaka, Osaka 560-8531 \\
$^{2}$National Institute of Advanced Industrial Science and Technology (AIST), Umezono, Tsukuba 305-8568 \\
$^{3}$JST, TRIP (Transformative Research-Project on Iron Pnictides), Chiyoda, Tokyo 102-0075 
}

\abst{
We report our extensive $^{57}$Fe-, $^{75}$As-, $^{139}$La-, and $^{1}$H-NMR studies of La$_{0.8}$Y$_{0.2}$FeAsO$_{1-y}$ (La$_{0.8}$Y$_{0.2}$1111) and  LaFeAsO$_{1-y}$H$_{x}$(La1111H), where doping yttrium (Y) and hydrogen (H) into optimally doped LaFeAsO$_{1-y}$ (La1111(OPT)) increases  $\sl{T}_{c}$=28 K to 34 and 32 K, respectively.  In the superconducting (SC) state, the measurements of nuclear-spin lattice-relaxation rate $1/T_1$ have revealed in terms of a multiple fully gapped $s_\pm$-wave model that the SC gap and $T_c$ in La$_{0.8}$Y$_{0.2}$1111 become larger than those in La1111(OPT) without any change in doping level. In La1111H, the SC gap and $T_c$ also increase slightly even though a decrease in carrier density and some disorders are significantly introduced. As a consequence, we suggest that the optimization of both the structural parameters and the carrier doping level to fill up the bands is crucial for increasing $T_c$ among these La1111-based compounds through the optimization of the Fermi surface topology. 
}

\kword{superconductivity, iron-based oxypnictide, LaFeAsO, NMR}

\begin{document}

\maketitle

\date{\today}

\begin{figure}[tbp]
\begin{center}
\includegraphics[width=0.9\linewidth]{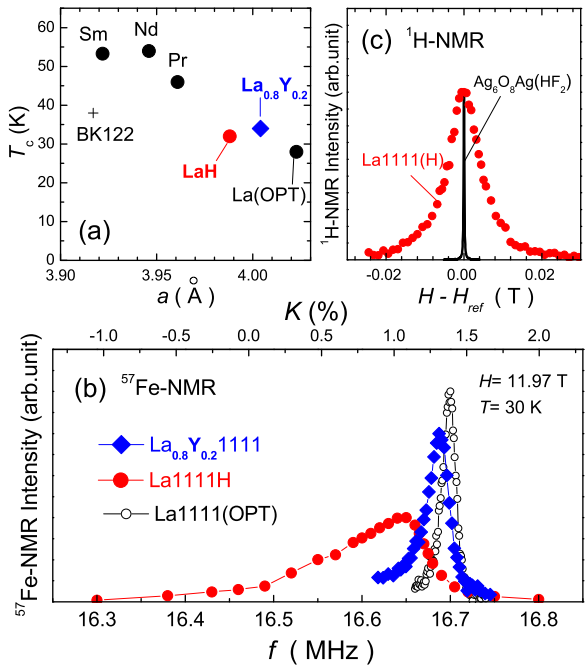}
\end{center}
\caption[]{\footnotesize (color online)
(a) Plot of $T_c$ vs $a$-axis length for La$_{0.8}$Y$_{0.2}$1111, La1111H, and $Ln$1111 systems~\cite{Shirage,Miyazawa1}. (b) Comparison of the $^{57}$Fe-NMR spectra of La$_{0.8}$Y$_{0.2}$1111 and La1111H at 30 K with that of La1111(OPT)~\cite{Terasaki}. (c) Comparison of the $^1$H-NMR spectrum in La1111H with that of stoichiometric compound Ag$_6$O$_8$Ag(HF$_2$) at $H\sim 3.916$ T and $T\sim$100 K. 
}
\label{spectra}
\end{figure}

Immediately after the discovery of superconductivity (SC) in the iron-oxypnictide LaFeAsO$_{1-x}$F$_x$ ($\sl{T}_{c}$~=~26 K)\cite{Kamihara2008}, it was reported that the replacement of La by other rare-earth ($Ln$) elements significantly increases the transition temperature $\sl{T}_{c}$ up to more than 50~K~\cite{Ren1,Kito,Ren2}. Lee {\it et al.} found that $T_{c}$ increases up to a maximum of 55 K when the FeAs$_{4}$ tetrahedron is transformed into a regular one~\cite{C.H.Lee}. Related to this, structural parameters such as the $a$-axis length~\cite{Ren2,Shirage,Miyazawa1} and height of pnictogen from the Fe plane~\cite{Mizuguchi} also exhibit an intimate correlation with $T_c$ in the $Ln$FeAsO($Ln$1111) system. Systematic measurements by spectroscopies have been performed extensively on $M$Fe$_{2}$As$_{2}$($M$122) and FeSe systems, probing the multiband character of their Fermi surfaces and the development of antiferromagnetic (AFM) spin fluctuations, but not sufficiently on $Ln$1111 systems because high-quality single crystals of sufficiently large size are not yet available\cite{Ishida}. In previous NMR studies of Nd1111 ($\sl{T}_{c}$=53~K) and Pr1111 ($\sl{T}_{c}$=47 K)~\cite{Yamashita,Jeglic}, $4f$-electron-derived magnetic fluctuations prevented us from deducing the normal-state properties and SC characteristics of FeAs layers. Under these situations, the reason why $T_c$ is highest in the $Ln$1111 system has not yet been addressed. Recently, it has been reported that $T_c$ can be increased by either Y or H substitution in the La1111 system without replacing magnetic rare-earth elements~\cite{Shirage,Tropeano,MiyazawaH}, in which the angle $\alpha$ of As-Fe-As bonding and the $a$-axis length approach those of the Nd1111 system with the highest $T_c$ to date (see Fig.~\ref{spectra}(a)).

In this Letter, we report the normal-state and SC characteristics of La$_{0.8}$Y$_{0.2}$FeAsO$_{1-y}$ ($T_c=$34 K) and LaFeAsO$_{1-y}$H$_{x}$ ($T_c=$32 K) determined using extensive NMR measurements of $^{57}$Fe, $^{75}$As, $^{139}$La, and $^{1}$H. We address the important correlation between the evolution of the electronic state caused by Y  and H substitutions and the optimization of the local structure of the FeAs$_4$ tetrahedron in these La1111 systems.


$^{57}$Fe-enriched polycrystalline samples of La$_{0.8}$Y$_{0.2}$FeAsO$_{0.7}$ and LaFeAsO$_{0.58}$H$_{0.58}$ each with a nominal composition were synthesized via a high-pressure synthesis technique~\cite{Kito,Shirage,MiyazawaH}. 
Although the oxygen and hydrogen contents of the samples differ from the nominal composition during the oxidation of the starting rare-earth elements, the X-ray diffraction measurements indicate that these samples are almost of a single phase. 
We hereafter denote these samples as La$_{0.8}$Y$_{0.2}$1111 and La1111H. The respective $\sl{T}_{c}$= 34 and 32 K for La$_{0.8}$Y$_{0.2}$1111 and La1111H were uniquely determined by a steep variation in susceptibility, being higher than $T_c$=28 K for La1111(OPT)~\cite{MukudaNQR,Terasaki}. As shown in Fig.~\ref{spectra}(a), which was reported in the literature~\cite{Ren2,Shirage,Miyazawa1}, note that the $a$-axis length of La1111H and La$_{0.8}$Y$_{0.2}$1111 is closer to the optimum value for reaching the maximum $T_c$ than that of optimally doped LaFeAsO$_{1-y}$(La1111(OPT)). The NMR measurements of $^{57}$Fe, $^{75}$As, $^{139}$La, and $^{1}$H were performed on most oriented powder samples of La$_{0.8}$Y$_{0.2}$1111 and La1111H. 
The nuclear spin-lattice relaxation rate ($1/T_1$) was measured in the field $H\perp c$ by the saturation-recovery method. 


Figure~\ref{spectra}(b) shows the $^{57}$Fe-NMR spectra obtained by sweeping a frequency ($f$) at a magnetic field $H=$~11.97~T at 30 K.  The $^{57}$Fe-NMR spectra become broader in La$_{0.8}$Y$_{0.2}$1111 and La1111H than in La1111(OPT)~\cite{Terasaki} as a result of the substitution of either Y or H.  In particular, an extremely broadened $^{57}$Fe-NMR spectral width of La1111H is almost independent of temperature, suggesting that H doping makes local magnetic states quite inhomogeneous through the distribution of uniform spin susceptibility and/or the hyperfine-coupling constant at the Fe site. 
By contrast, the $^{57}$Fe-NMR spectral width of La$_{0.8}$Y$_{0.2}$1111 is significantly narrower than that of La1111H. This suggests that the Y$^{3+}$ substitution for La$^{3+}$ introduces fewer disorders than H-doping. 
The $^1$H-NMR spectrum for La1111H is presented in Fig.~\ref{spectra}(c). Note that the $^1$H-NMR spectral width of approximately 90 Oe is much broader than those of stoichiometric compounds including H ions located at a regular crystallographic site, for instance, 2.6 Oe in Ag$_6$O$_8$Ag(HF$_2$). This may be due to the wide distribution of the transferred hyperfine fields at the H site induced by Fe-spin polarization. 
Since the $^1$H-NMR spectral shape is symmetric, most of the H ions may occupy a single site within the LaFeAsO structure. 
\begin{figure}[htbp]
\begin{center}
\includegraphics[width=0.9\linewidth]{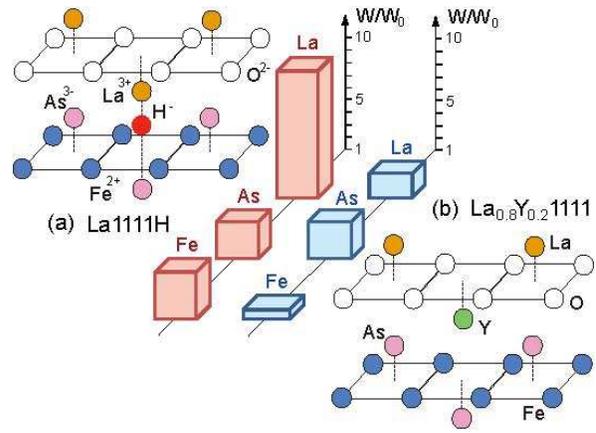}
\end{center}
\caption[]{\footnotesize (color online)
Site dependences of the NMR spectral width ($W$) at 40 K of $^{57}$Fe, $^{75}$As, and $^{139}$La for (a) La1111H and (b) La$_{0.8}$Y$_{0.2}$1111. Each $W$ is normalized by $W_0$ defined by those of La1111(OPT). In La1111H (La$_{0.8}$Y$_{0.2}$1111), each $W/W_0$ at the Fe and La sites is larger (smaller) than that at the As site, suggesting that the H ions predominantly occupy an interstitial site between the La and Fe sites. Taking the decrease in the carrier density of La1111H into account, the doped H ions will exist as negatively charged H$^{-}$ ions.
}
\label{fig:linewidth}
\end{figure}

In order to shed further light on the local disorder introduced by substitution, we compare the  site dependences of the NMR spectral widths ($W$) at 40 K measured by $^{57}$Fe-, $^{75}$As-, and $^{139}$La-NMR in La$_{0.8}$Y$_{0.2}$1111 and La1111H, as shown in Fig.~\ref{fig:linewidth}. 
Here, each $W$ is normalized by $W_0$ defined by that of La1111(OPT). In La1111H (La$_{0.8}$Y$_{0.2}$1111), the ratio of the linewidth $W/W_0$ at the Fe and La sites is larger (smaller) than that at the As site, suggesting that the H ions occupy an interstitial site between the La and Fe sites, which induces local disorder especially at Fe and La sites. 
Taking the decrease in the carrier density of La1111H into account, to be discussed later, 
the doped H ions will exist as negatively charged H$^{-}$ ions, which may be stable at the interstitial site surrounded by positively charged La$^{3+}$ and Fe$^{\sim 2+}$ ions, as indicated in the inset of Fig.~\ref{fig:linewidth}(a). 
\begin{figure}[htbp]
\begin{center}
\includegraphics[width=1\linewidth]{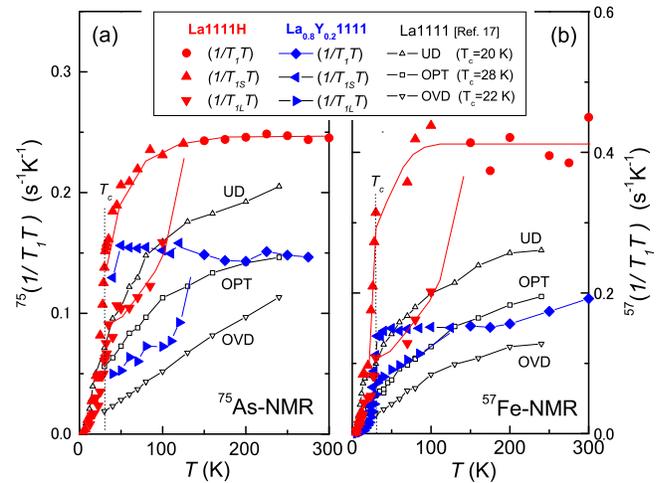}
\end{center}
\caption[]{\footnotesize (color online)
$T$ dependences of $1/T_{1}T$s of (a)$^{75}$As- and (b) $^{57}$Fe-NMR for La$_{0.8}$Y$_{0.2}$1111 and La1111H. The comparison with those of La1111 compounds~\cite{MukudaFe2} reveals that the electron doping level of La$_{0.8}$Y$_{0.2}$1111 is close to that of La1111(OPT)~\cite{MukudaFe2,Nakai2}, but that La1111H is in the underdoped regime.
}
\label{fig:FeNMRT1}
\end{figure}

Figure~\ref{fig:FeNMRT1}(a) shows the $T$ dependences of $^{75}(1/T_1T)$ for La$_{0.8}$Y$_{0.2}$1111 and La1111H. In both compounds, $^{75}(1/T_1)$ can be determined by a single $T_1$ component above $\sim$150 K, indicating that the electronic state seems to be almost uniform over the sample. At $T$ range below $\sim$150 K, however, $^{75}T_1$ exhibits an apparent distribution in association with the substitution of either Y or H. 
The short component $^{75}T_{1S}$ and the long component $^{75}T_{1L}$, determined by the same method in ref. 16, are plotted in the figure. 
In La$_{0.8}$Y$_{0.2}$1111, the $^{75}(1/T_{1S}T)$ shows a nearly constant behavior above $T_c$, whereas $^{75}(1/T_{1L}T)$ shows a gradual decrease upon cooling, which resembles the $^{75}$As-NMR result of La1111(OPT)~\cite{MukudaFe2}. It was reported that the $^{75}(1/T_1T)$ at high $T$ decreases markedly as the doping level of electron carriers increases in the La1111 system~\cite{MukudaFe2,Nakai2}, which was corroborated by $^{57}(1/T_1T)$, as indicated in Fig.~\ref{fig:FeNMRT1}(b).  Here, we note that the $^{57,75}(1/T_{1}T)$ at 250 K for La$_{0.8}$Y$_{0.2}$1111 is comparable to that for La1111(OPT). This reveals that the electron doping level of La$_{0.8}$Y$_{0.2}$1111 is close to that of La1111(OPT), confirming that the Y$^{3+}$ substitution for La$^{3+}$ does not change the doping level. 
By contrast, in  La1111H, it is remarkable that the $1/T_{1}T$s of both $^{57}$Fe- and $^{75}$As-NMR at 250 K are markedly larger than those of La1111(OPT), the latter being comparable to $^{75}(1/T_{1}T)$ for the underdoped LaFeAsO$_{0.93}$F$_{0.07}$ ($T_c=$ 22.5 K)~\cite{Nakai2}. This implies that La1111H is in an underdoped regime. When noted that the doped H atoms are located at the interstitial site surrounded by positively charged Fe$^{\sim 2+}$ and La$^{3+}$ sites (see Fig.\ref{fig:linewidth}(a)), it is anticipated that the doped H atoms will exist as H$^-$ ions to reduce the doping level.
Furthermore, the significant increase in $1/T_1T$ upon cooling was not observed in both compounds even though the Knight shift for La$_{1-x}$Y$_{x}$1111 is almost constant against $T$\cite{Yamashita}, suggesting that strong AFM spin fluctuations do not develop in these compounds.  

\begin{figure}[htbp]
\begin{center}
\includegraphics[width=0.8\linewidth]{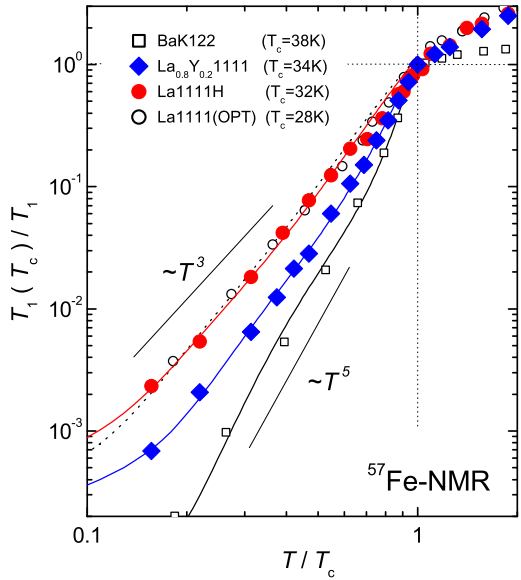}
\end{center}
\caption[]{\footnotesize (color online)
$T$ dependences of $^{57}$Fe-$(T_1(T_c)/T_1)$ normalized  at $T_c$ in the SC state of La$_{0.8}$Y$_{0.2}$1111 and La1111H, along  with the results reported for BaK122 and La1111(OPT) \cite{Yashima}. The multiple fully gapped $s_\pm$-wave model allows us to deduce the parameters for all the compounds listed in Table \ref{table1} through the fitting of the $T$ dependence of $^{57}$Fe-$(T_1(T_c)/T_1)$. Here, the $T_{1L}$ components are plotted for La$_{0.8}$Y$_{0.2}$1111 and La1111H, since both $T_1$-components show the same $T$ dependence when normalized by the values at $T_c$. 
}
\label{fig:FeNMRT1SC}
\end{figure}

Next, we address the SC characteristics of these compounds. Figure \ref{fig:FeNMRT1SC} shows the $\sl{T}$ dependence of $^{57}(T_1(T_c)/\sl{T}_{1})$ normalized at their $T_c$s. Since the $T$ dependences of both $T_{1S}$ and $T_{1L}$ are almost the same, $^{57}(T_{1L}(T_c)/\sl{T}_{1L})$ is presented in the figure.  The $^{57}(T_{1L}(T_c)/\sl{T}_{1L})$ of La$_{0.8}$Y$_{0.2}$1111 decreases similarly to $\sim T^4$ upon cooling below $\sl{T}_{c}$, which differs from either the $\sim\sl{T}^{3}$ in La1111(OPT)~\cite{Terasaki} or the $\sim\sl{T}^{5}$ in Ba$_{0.6}$K$_{0.4}$Fe$_{2}$As$_{2}$(BaK122)~\cite{Yashima}. 
This emphasizes that a common power law like the $T$ dependence of $1/T_1$ is not evident among Fe-based superconductors. These results are in contrast to the behavior of $1/T_1\sim T^3$ commonly observed in high-$T_c$ cuprates, which are $d$-wave superconductors with line-node gaps. 
Meanwhile, it was shown in the literature that the relaxation behaviors of both the $\sim\sl{T}^{3}$ in La1111(OPT)~\cite{Terasaki,Nagai} and the $\sim\sl{T}^{5}$ in Ba$_{0.6}$K$_{0.4}$Fe$_{2}$As$_{2}$ (BaK122)~\cite{Yashima} are consistently reproduced in terms of the multiple fully gapped $s_\pm$-wave model\cite{Nagai}. This model is also applicable for understanding the SC characteristics of La$_{0.8}$Y$_{0.2}$1111 and La1111H as follows. 
According to Model B in the literature~\cite{Yashima}, we assume that respective Fermi surfaces (FS1 and FS2) have isotropic gaps as $\Delta^{FS1}\equiv \Delta_L$ and $\Delta^{FS2}\equiv \Delta_S$, and the fraction of the density of states (DOS) at FS1 is taken as $N_{FS1}/(N_{FS1}+N_{FS2})$(=0.7)\cite{model}.
Furthermore, the coherence factor is neglected on the assumption that the interband scattering between the sign reversal gaps becomes dominant for the relaxation process in these compounds. In fact, the $T$ dependence of $^{57}(T_1(T_c)/\sl{T}_{1})$ of La$_{0.8}$Y$_{0.2}$1111 is consistently reproduced with the parameters $2\Delta_L/k_BT_c$ = 6.9$(\Delta_{S}/\Delta_L$=0.35) and the smearing factor $\eta=0.04\Delta_L$\cite{model}, as is shown by the solid line in Fig.~\ref{fig:FeNMRT1SC}.  A damping effect of quasiparticles due to impurity scattering introduced by the Y substitution can be deduced from $\eta$, which was actually larger than $\eta_0$ in La1111(OPT)(see Table~\ref{table1}). 
Notably, $2\Delta_L/k_BT_c$= 6.9 in La$_{0.8}$Y$_{0.2}$1111 is larger than the $2\Delta_L/k_BT_c$  4.4 in La1111(OPT), revealing that the Y substitution into La1111 increases $T_c$ up to 34 K in association with a strong-coupling effect to mediate the Cooper pairs. Note that the strong-coupling effect enhances $T_c$ despite the disorder introduced by Y substitution. 

\begin{table}[htbp]
\centering
\caption[]{\footnotesize 
Evolutions of the SC gap ($\Delta_L$) and smearing factor ($\eta$) obtained from the analyses of $^{57}$Fe-$(T_1(T_c)/T_1)$  assuming the multiple fully gapped $s_\pm$ wave model (Model B) that was applied to BK122 and La1111(OPT)~\cite{Yashima}(see text). Here, $\Delta_L$ represents the larger of the two full gaps. $\eta/\eta_0$ represents a damping effect of quasiparticles additionally introduced by Y or H substitution, where $\eta_0$ is defined as that of La1111(OPT).}
\begin{tabular}{lcccccc}
\hline 
                  & $T_c$ & $a$  & $\alpha$($^\circ$)$^\dagger$ & $2\Delta_L$ & $\eta/\Delta_L$ & $\eta/\eta_0$ \\
                  & (K)  & (\AA) &                    & $/k_BT_c$   &                 &      \\
\hline
BaK122\cite{Yashima} & 38 & 3.914 & 109.7\cite{Rotter} &  9.4       & 0.015         & 0.64 \\
La$_{0.8}$Y$_{0.2}$1111     & 34 & 4.004 & 112.3 &  6.9             & 0.04          & 1.3  \\ 
La1111H                     & 32 & 3.989 & 111.7 &  $\sim$4.7       & $\sim$0.06    & $\sim$1.3  \\    
La1111(OPT)\cite{Yashima}   & 28 & 4.023 & 113.2 &  4.4             & 0.05          & 1    \\
\hline
\end{tabular}
\footnotesize{$\dagger$) The angle $\alpha$ of As-Fe-As bonding is evaluated from the $a$-axis length at room $T$ and a fixed bond length of Fe-As $\sim$ 2.41\AA, which holds empirically in La1111 system~\cite{C.H.Lee}.}
\label{table1}
\end{table} 

As for La1111H, it is possible that $^{57}(T_{1L}(T_c)/{T}_{1L})$ is also reproduced roughly in terms of Model B with the parameters $2\Delta_L/k_BT_c\sim 4.7$($\Delta_{S}/\Delta_L$=0.35) and $\eta\sim 0.06\Delta_L$, similarly to those in La1111(OPT), as shown by the solid line in Fig.~\ref{fig:FeNMRT1SC}. 
Despite the fact that the electron carrier density is lower than that in La1111(OPT) and the disorder is heavily introduced into the Fe site, which is deduced from $\eta/\eta_0\sim1.3$, $T_c$ is enhanced up to 32 K. As a result, the reason why $T_c$ increases in La1111H may be closely related to the slight increase in the SC gap as well.  

All the parameters listed in Table~\ref{table1} are obtained from the analyses of the $T_1$ results based on the two fully gapped $s_\pm$ wave model(Model B). In the case of BaK122, which has the largest gap of $2\Delta_L/k_BT_c$ = 9.4, the development of AFM spin fluctuations is argued to be the origin of the strong-coupling effect, making the SC gap quite large~\cite{Yashima}. 
By contrast, the $T_1$ measurements in the normal state do not point to the development of AFM spin fluctuations in La$_{0.8}$Y$_{0.2}$1111 or other La1111 systems~\cite{MukudaFe2}. 
The reason why $2\Delta_L/k_BT_c=6.9$ is larger in La$_{0.8}$Y$_{0.2}$1111 than in La1111(OPT) is not primarily ascribed to AFM spin fluctuations at low energies. We suggest that the structural parameter $\alpha$=109.7$^\circ$ for BaK122~\cite{Rotter} is the same as that of the regular tetrahedron, whereas the $\alpha$=112.3$^\circ$ for La$_{0.8}$Y$_{0.2}$1111 is larger. This suggests that the SC energy gap increases as $\alpha$ approaches $\alpha$=109.47$^\circ$ for a  regular tetrahedron. Furthermore, note that $T_c$ increases as the $a$-axis length decreases in going from La1111(OPT) to BaK122, as shown in Table~\ref{table1}. 
Here, we should comment on why the $T_c$ in La1111H is lower than that in La$_{0.8}$Y$_{0.2}$1111, despite the structural parameters such as the $a$-axis length and the angle $\alpha$ being closer to those of BK122 than to those of La$_{0.8}$Y$_{0.2}$1111; This may be because the electron doping level in La1111H is lower than that in La$_{0.8}$Y$_{0.2}$1111, in addition to some disorder effect. 
In this context, the optimization of both the structural parameters and the carrier doping level to fill up the bands is crucial for increasing their $T_c$ through the optimization of the Fermi surface topology. 


In summary, the systematic NMR measurements of La$_{0.8}$Y$_{0.2}$FeAsO$_{1-y}$  ($\sl{T}_{c}$=34 K) and LaFeAsO$_{1-y}$H$_{x}$ ($\sl{T}_{c}$=32 K) have revealed that the Y$^{3+}$ substitution does not change the doping level in La$_{0.8}$Y$_{0.2}$1111, whereas H-doping decreases the carrier density in La1111H. The $1/T_1$ results in the SC state for both compounds are consistently interpreted in terms of the multiple fully gapped $s_\pm$-wave model. Thus, it is highlighted that the SC gap and $T_c$ in La$_{0.8}$Y$_{0.2}$1111 become larger than those in La1111(OPT) without any change in doping level. Furthermore, the $T_c$ and SC gap in La1111H slightly increases even though the decrease in carrier density and some disorders are heavily introduced.  
We suggest that the primary reason why $T_c$ is increased in these La-based 1111 compounds is neither the change in doping level nor the development of AFM spin fluctuations, but the structural parameters approaching their optimum values to increase $T_c$ for the bond angle $\alpha$ of the FeAs$_4$ tetrahedron and the $a$-axis length. Systematic spectroscopies of the SC properties of $Ln$1111 systems with the highest $T_c$ of more than $50$ K are highly desired in the future in order to clarify the Fermi surface topology, the antiferromagnetic spin fluctuations in the normal state, and their relevance to SC gap structures.


We thank C. H. Lee for valuable discussions. This work was supported by a Grant-in-Aid for Specially Promoted Research (20001004) and by the Global COE Program (Core Research and Engineering of Advanced Materials-Interdisciplinary Education Center for Materials Science) from the Ministry of Education, Culture, Sports, Science and Technology (MEXT), Japan.



\begin{thebibliography}{99} 

\bibitem{Kamihara2008} Y. Kamihara, T. Watanabe, M. Hirano, and H. Hosono: J. Am. Chem. Soc. {\bf 130} (2008) 3296.
\bibitem{Ren1} Z. A. Ren, W. Lu, J. Yang, W. Yi, X. L. Shen, Z. C. Li, G. C. Che, X. L. Dong, L. L. Sun, F. Zhou, and Z. X. Zhao: Chin. Phys. Lett. {\bf 25} (2008) 2215.
\bibitem{Kito} H. Kito, H. Eisaki, and A. Iyo: J. Phys. Soc. Jpn. {\bf 77} (2008) 063707.
\bibitem{Ren2} Z. A. Ren, G. C. Che, X. L. Dong, J. Yang, W. Lu, W. Yi, X. L. Shen, Z. C. Li, L. L. Sun, F. Zhou, and Z. X. Zhao: Europhys. Lett. {\bf 83} (2008) 17002.
\bibitem{C.H.Lee} C. H. Lee, H. Eisaki, H. Kito, M. T. Fernandez-Diaz, T. Ito, K. Kihou, H. Matsushita, M. Braden, and K. Yamada: J. Phys. Soc. Jpn. {\bf 77} (2008) 083704.
\bibitem{Shirage} P. M. Shirage, K. Miyazawa, H. Kito, H. Eisaki, and A. Iyo: Phys. Rev. B  {\bf 78} (2008) 172503.
\bibitem{Miyazawa1} K. Miyazawa, K. Kihou, P. M. Shirage, C.-H Lee, H. Kito, H. Eisaki, and A. Iyo: J. Phys. Soc. Jpn. {\bf 78} (2009) 034712.
\bibitem{Mizuguchi} Y. Mizuguchi, Y. Hara, K. Deguchi, S. Tsuda, T. Yamaguchi, K. Takeda, H. Kotegawa, H. Tou, and Y. Takano: Supercond. Sci. Technol. {\bf 23} (2010) 054013.
\bibitem{Ishida} K. Ishida, Y. Nakai, and H. Hosono: J. Phys. Soc. Jpn. {\bf 78} (2009) 062001.
\bibitem{Yamashita} H. Yamashita, M. Yashima, H. Mukuda, Y. Kitaoka, P. M. Shirage, and A. Iyo: Physica C (2009), doi:10.1016/j.physc.2009.11.125.
\bibitem{Jeglic} P. Jegli$\check{c}$, J.-W. G. Bos, A. Zorko, M. Brunelli, K. Koch, H. Rosner, S. Margadonna, and D. Ar$\check{c}$on: Phys. Rev. B. {\bf 79} (2009) 094515.
\bibitem{Tropeano} M. Tropeano, C. Fanciulli, F. Canepa, M. R. Cimberle, C. Ferdeghini, G. Lamura, A. Martinelli, M. Putti, M. Vignolo, and A. Palenzona: Phys. Rev. B  {\bf 79} (2009) 174523. 
\bibitem{MiyazawaH} K. Miyazawa, S. Ishida, K. Kihou, P. M. Shirage, M. Nakajima, C. H. Lee, H. Kito, Y. Tomioka, T. Ito, H. Eisaki, H. Yamashita, H. Mukuda, K. Tokiwa, S. Uchida, and A. Iyo: Appl. Phys. Lett. {\bf 96} (2010) 072514.
\bibitem{MukudaNQR} H. Mukuda, N. Terasaki, H. Kinouchi, M. Yashima, Y. Kitaoka, S. Suzuki, S. Miyasaka, S. Tajima, K. Miyazawa, P. M. Shirage, H. Kito, H. Eisaki, and A. Iyo: J. Phys. Soc. Jpn. {\bf 77} (2008) 093704. 
\bibitem{Terasaki} N. Terasaki, H. Mukuda, M. Yashima, Y. Kitaoka, K. Miyazawa, P.M. Shirage, H. Kito, H. Eisaki, and A. Iyo: J. Phys. Soc. Jpn. {\bf 78} (2009) 013701.
\bibitem{MukudaFe2}  H. Mukuda, N. Terasaki, N. Tamura, H. Kinouchi, M. Yashima, Y. Kitaoka, K. Miyazawa, P. M. Shirage, S. Suzuki, S. Miyasaka, S. Tajima, H. Kito, H. Eisaki, and A. Iyo: J. Phys. Soc. Jpn. {\bf 78} (2009) 084717.
\bibitem{Nakai2} Y. Nakai, S. Kitagawa, K. Ishida, Y. Kamihara, M. Hirano, and H. Hosono: New J. Phys. {\bf 11} (2009) 045004.
\bibitem{Yashima} M. Yashima, H. Nishimura, H. Mukuda, Y. Kitaoka, K. Miyazawa, P. M. Shirage, K. Kiho, H. Kito, H. Eisaki, and A. Iyo: J. Phys. Soc. Jpn. {\bf 78} (2009) 103702.
\bibitem{Nagai} Y. Nagai, N. Hayashi, N. Nakai, H. Nakamura, M. Okumura, and M. Machida: New J. Phys. {\bf 10} (2008) 103026.
\bibitem{model} We note that both Models A and B in the literature\cite{Yashima} can reproduce the present results with similar parameters, although only the results based on Model B are shown. 
\bibitem{Rotter} M. Rotter, M. Tegel, and D. Johrendt: Phys. Rev. Lett. {\bf 101} (2008) 107006.

\end{thebibliography}
\end{document}